\documentclass[aps,prl,twocolumn, showpacs]{revtex4-1}
\pdfoutput=1
\usepackage[latin9]{inputenc}
\usepackage{amsmath}
\usepackage{graphicx}
\usepackage{amssymb}
\usepackage{esint}

\begin{document}

\title{Coupled multimode optomechanics in the microwave regime}

\author{Georg Heinrich}

\affiliation{Arnold Sommerfeld Center for Theoretical Physics, Center for NanoScience
and Department of Physics, Ludwig-Maximilians-Universität München,
Theresienstr. 37, D-80333 München, Germany\\
Institut f\"ur Theoretische Physik, Universit\"at Erlangen-N\"urnberg, Staudtstr. 7, 91058 Erlangen, Germany}

\author{Florian Marquardt}

\affiliation{Arnold Sommerfeld Center for Theoretical Physics, Center for NanoScience
and Department of Physics, Ludwig-Maximilians-Universität München,
Theresienstr. 37, D-80333 München, Germany\\
Institut f\"ur Theoretische Physik, Universit\"at Erlangen-N\"urnberg, Staudtstr. 7, 91058 Erlangen, Germany}

\date{\today}

\begin{abstract}
The motion of micro- and nanomechanical resonators can be coupled
to electromagnetic fields. This allows to explore the mutual interaction
and introduces new means to manipulate and control both light and
mechanical motion. Such optomechanical systems have recently been
implemented in nanoelectromechanical systems involving a nanomechanical beam
coupled to a superconducting microwave resonator. Here, we propose
optomechanical systems that involve multiple, coupled microwave resonators. In
contrast to similar systems in the optical realm, the coupling frequency
governing photon exchange between microwave modes is naturally comparable
to typical mechanical frequencies. For instance this enables new ways
to manipulate the microwave field, such as mechanically driving coherent
photon dynamics between different modes. In particular we investigate
two setups where the electromagnetic field is coupled either linearly
or quadratically to the displacement of a nanomechanical beam.
The latter scheme allows to perform QND Fock state detection. For
experimentally realistic parameters we predict the possibility to measure
an individual quantum jump from the mechanical ground state to the first excited state.
\end{abstract}


\maketitle

\textbf{Introduction.} - Significant interest in the interaction and
dynamics of systems comprising micro- and nanomechanical resonators
coupled to electromagnetic fields, as well as the prospect to eventually
measure and control the quantum regime of mechanical motion, has stimulated
the rapidly evolving field of optomechanics (see \cite{Marquardt2009Optomechanics}
for a recent review). In the standard setup, the light field, stored
inside an optical cavity, exerts a radiation pressure force on a movable
end-mirror whose motion changes the cavity frequency and thus acts
back on the photon dynamics. This way, the photon number inside the
optical mode is linearly coupled to the displacement of a mechanical
object. Beyond the standard approach, new developments have introduced
optical setups with multiple coupled light and vibrational modes pointing
the way towards integrated optomechanical circuits \cite{ThompsonStrong-dispersi,Li2008Harnessing-opti,Eichenfield2009A-picogram--and,Eichenfield2009Optomechanical-,Anetsberger2009Near-field-cavi}.
These systems allow to study elaborate interactions between mechanical
motion and light such as mechanically driven coherent photon dynamics
that introduces the whole realm of driven two- and multi-level systems
to the field of optomechanics \cite{Heinrich2010Photon-shuttle:}.
Coupled multimode setups furthermore allow to increase
measurement sensitivity \cite{Dobrindt2010Theoretical-Ana} and enable fundamentally different
coupling schemes. Accordingly, recent experiments achieved coupling
the photon number to the square and quadruple of mechanical displacement
\cite{ThompsonStrong-dispersi,Sankey2010Strong-and-Tuna}. Such different
coupling schemes are needed, for instance, to afford quantum non-demolition
(QND) Fock state detection of a mechanical resonator \cite{ThompsonStrong-dispersi, Braginsky1980Quantum-Nondemo,Braginsky1992Quantum-Measure,Jayich2008Dispersive-opto}.

Besides optics, recent progress has made it possible to realize optomechanical
systems in the microwave regime \cite{Regal2008Measuring-nanom,Hertzberg2010Back-action-eva}.
In this case the optical cavity is replaced by a superconducting microwave
resonator whose central conductor capacitively couples to the motion
of a nanomechanical beam. This optomechanical approach constitutes
a new path to perform on-chip experiments measuring and manipulating
nanomechanical motion that adds to electrical concepts using single
electron transistors \cite{2003Natur.424..291K,LaHaye2004Approaching-the,Naik2006Cooling-a-nanom},
superconducting quantum interference devices \cite{Etaki2008Motion-detectio,Buks2008Quantum-nondemo},
driven RF circuits \cite{Brown2007Passive-Cooling} or a Cooper-pair-box
\cite{LaHaye2009Nanomechanical-,O/Connell2010Quantum-ground-}. One advantage of on-chip optomechanics
is to use standard bulk refrigerator techniques in addition to laser
cooling schemes \cite{Teufel2008Dynamical-Backa,Teufel2008Prospects-for-c}.
This recently enabled cooling a single vibrational mode close to the
quantum mechanical ground state \cite{Rocheleau2010Preparation-and}.
Furthermore, nonlinear circuit elements can be integrated. This afforded
ultra-sensitive displacement measurements with measurement imprecision below
that at the standard quantum limit \cite{Teufel2009Nanomechanical-}.

Here we go beyond single mode systems, that have been considered for
optomechanics in the microwave regime so far, and propose setups
with coupled microwave resonators. For these systems, the coupling frequency between
microwave modes turns out to be comparable to the mechanical frequency. This has
several implications both for the classical and the prospective quantum
regime. As an example, focusing on the currently in experiments accessible regime of classical motion, we demonstrate how this allows to manipulate the
microwave field in terms of mechanical driving.

\begin{figure}
\includegraphics[width=1\columnwidth]{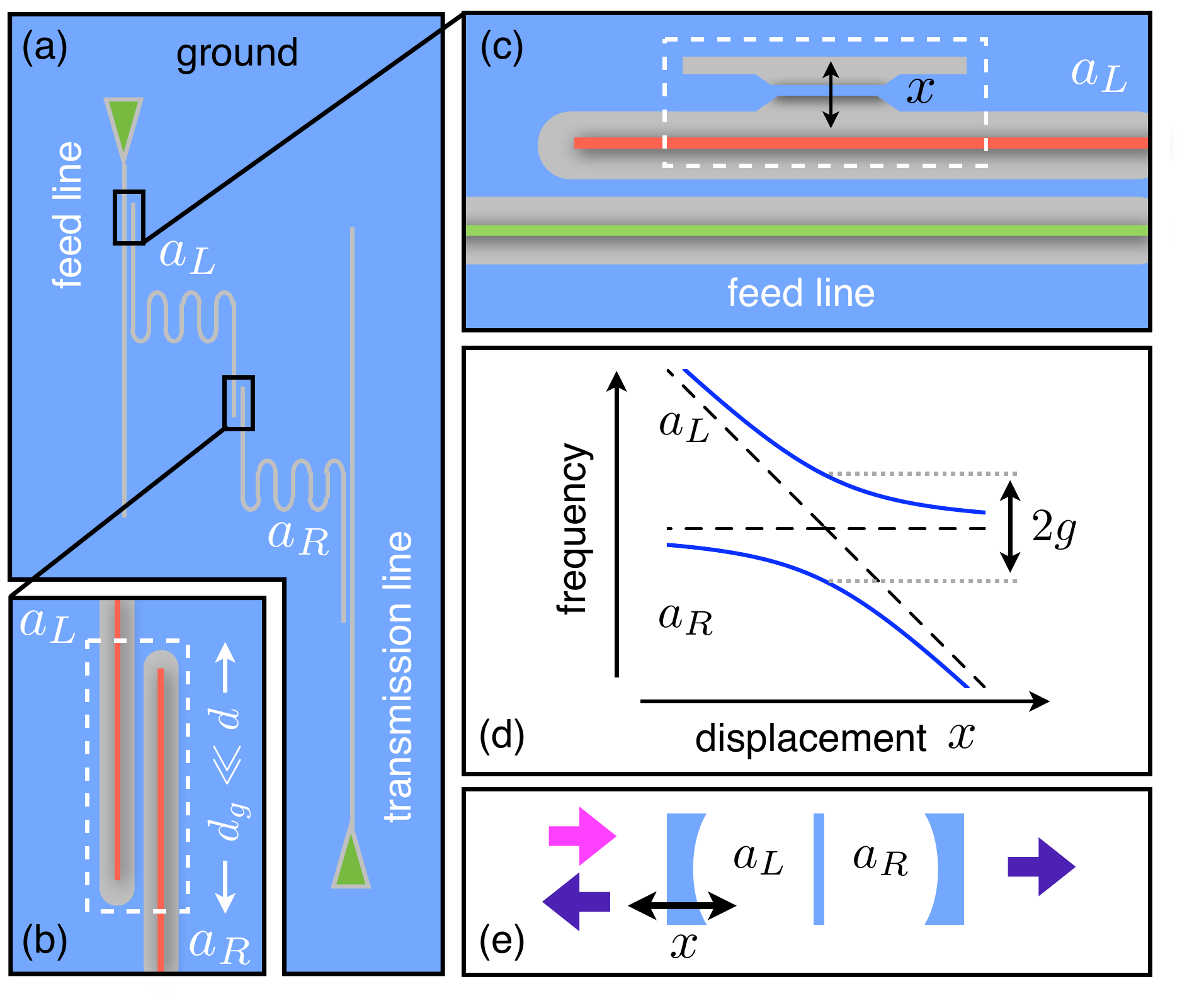}

\caption{\label{fig:Setup_linear}Schematic device geometry for two superconducting
microwave resonators $a_{L}$, $a_{R}$ with a nanomechanical beam
coupled to $a_{L}$. (a) The two resonators (each of length $d$)
are coupled to external feed and transmission lines (green). (b) The
central conductors of $a_{L}$ and $a_{R}$ (red) capacitively couple
due to a small region of length $d_{g}$ where the resonators adjoin.
(c) At the other end of $a_{L}$ a small mechanical beam, connected
to ground (blue), is placed. Its displacement $x$ affects the line
capacitance of resonator $a_{L}$ changing its resonance frequency.
(d) System's resonance frequency as function of displacement: the
beam's displacement $x$ linearly changes the bare mode frequency
of $a_{L}$ while the one of $a_{R}$ is unaffected (dashed). Due
to the coupling $g$ between modes there is an avoided crossing $2g$
in the eigenfrequencies (blue). (e) Analogous optical setup: a static, dielectic membrane, placed inside a cavity with a movable mirror, couples two separate optical modes $a_{L}$, $a_{R}$.}

\end{figure}

\textbf{Two-resonators setup with linear mechanical coupling.} - We
consider the coplanar device geometries depicted in Fig.~\ref{fig:Setup_linear}a
with two identical superconducting microwave resonators $a_L$, $a_R$.%
The central conductors of $a_{L}$ and $a_{R}$
are assumed to adjoin for a length $d_{g}$ that is much
smaller than the total wave guides' length $d$ {[}Fig.~\ref{fig:Setup_linear}b{]}.
A nanomechanical beam, connected to the ground plane, is placed at
the other end of $a_{L}$ {[}Fig.~\ref{fig:Setup_linear}c{]}. Its
motion in terms of displacement $x$ changes the line capacitance
$c$ (capacitance per unit length) between the central conductor and
the ground plane in a small region. In the following, we will derive
and discuss the Hamiltonian for the system depicted
in Fig.~\ref{fig:Setup_linear} starting from a single microwave resonator whose
line capacitance is changed due to the motion of a mechanical beam [Fig.~\ref{fig:Setup_linear}b]
\cite{Regal2008Measuring-nanom,Teufel2008Dynamical-Backa,Rocheleau2010Preparation-and}.

\textit{Single MW resonator coupled to a nanomechanical beam.} - We
concentrate on one of the microwave resonators in Fig.~\ref{fig:Setup_linear}.
The Lagrangian of an electric circuit can be conveniently expressed
in terms of a flux variable $\phi(x,t)\equiv\int_{-\infty}^{t}d\tau\, V(x,\tau),$
where $V(x,t)=\partial_{t}\phi(x,t)$ is the voltage on the transmission
line at position $x$ and time $t$, see for instance \cite{Devoret1995Quantum-Fluctua}.
For a finite length microwave resonator with line inductance $l$ and
line capacitance $c$ (both per unit length), the corresponding Euler-Lagrange
equation yields a wave equation with speed $v=\sqrt{1/lc}$. Using
appropriate boundary conditions,
$\phi(x,t)$ can be expanded into normal modes $\phi_k$, see \cite{Blais2004Cavity-quantum-}.
The Lagrangian reads $\mathcal{L}=\sum_{k}[\frac{c}{2}\dot{\phi}_{k}^{2}+\frac{1}{2l}\left(\frac{k\pi}{d}\right)^{2}\phi_{k}^{2}]$ and
the microwave resonator possesses resonance frequencies $\omega_{k}=k\pi v/d$.
The Hamiltonian is obtained by Legendre transformation using the canonically
conjugated momentum $\pi_{k}=c\dot{\phi}_{k}$. Quantizing the system
by introducing creation and annihilation operators $a_{k}^{\dagger}$,
$a_{k}$ for the individual modes $k$, with $[a_{k},\, a_{k'}^{\dagger}]=\delta_{k,k'}$,
the Hamiltonian is a sum of harmonic oscillators, $\sum_{k}\hbar\omega_{k}a_{k}^{\dagger}a_{k}$,
where we neglected the vacuum energy.

For the resonator $a_{L}$ in Fig.~\ref{fig:Setup_linear}, the motion
of the mechanical beam will change the line capacitance $c$ at the
end of the wave guide. If we denote the total capacitance between the central
conductor and the beam by $C_{b}$ and define the optomechanical frequency
pull per displacement, $\epsilon_{k}=-\partial\omega_{k}/\partial x$,
we find\begin{eqnarray}
\epsilon_{k} & = & \frac{\partial C_{b}}{\partial x}\cdot Z\cdot\omega_{k}^{2}/\left(2\pi k\right),\label{eq:def_epsilon}\end{eqnarray}
where we used the line impedance $Z=\sqrt{l/c}$. Up to linear order
in $x$, the Hamiltonian for a single microwave resonator coupled
to a nanomechanical beam reads $\mathcal{H}=\sum_{k}\hbar\left(\omega_{k,0}-\epsilon_{k}x\right)a_{k}^{\dagger}a_{k}$.

\begin{figure}
\begin{centering}
\includegraphics[width=1\columnwidth]{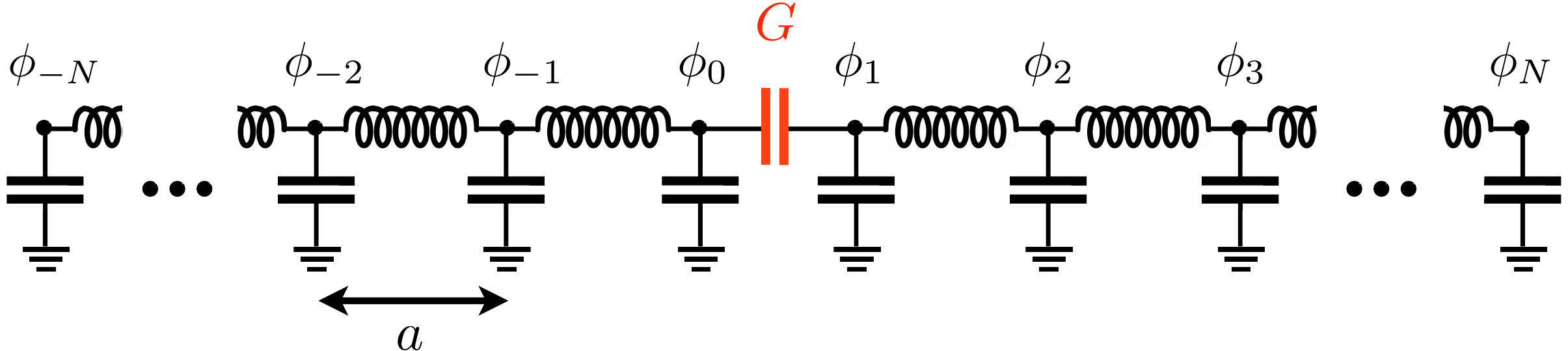}
\par\end{centering}

\caption{Discretized circuit diagram for the system depicted in Fig.~\ref{fig:Setup_linear}
without nanomechanical beam. The two central conductors are coupled
via a total capacitance $G$. $l$ and $c$ are the line inductance
and capacitance, respectively. $a$ is the spacing between discrete
points. \label{fig:Circuit-diagram}}

\end{figure}

\textit{Coupled MW resonators.} - We now turn to the coupled resonator
setup in Fig.~\ref{fig:Setup_linear}. As the length $d_{g}$ of
the region where $a_{L}$ and $a_{R}$ adjoin is much smaller than
the resonators' length $d$, the capacitive coupling between both
central conductors can be considered in terms of a constant, total
capacitance $G$. We
briefly switch to a discretized description. If we neglect the motion
of the nanomechanical beam the circuit diagram looks as depicted in
Fig.~\ref{fig:Circuit-diagram}, where the flux variable $\phi(x,t)$
is discretized as $\phi_{n}(t)$ at position $x=na$, with $a$ being
the constant spacing between nodes. %
For each individual node, an equation of motion can be written down \cite{Devoret1995Quantum-Fluctua}.
From this, taking into account appropriate boundary conditions for $a\rightarrow0$, the
continuous version of the corresponding Lagrangian is found to be
\begin{eqnarray}
\mathcal{L} & = & \sum_{k}\left[\frac{c}{2}\dot{\phi}_{L,k}^{2}-\frac{1}{2l}\left(\frac{k\pi}{d}\right)^{2}\phi_{L,k}^{2}\right]\nonumber \\
 & + & \sum_{k}\left[\frac{c}{2}\dot{\phi}_{R,k}^{2}-\frac{1}{2l}\left(\frac{k\pi}{d}\right)^{2}\phi_{R,k}^{2}\right]\nonumber \\
 & + & \frac{G}{d}\left(\sum_{k}\dot{\phi}_{R,k}-\sum_{k}\dot{\phi}_{L,k}\right)^{2},\label{eq:Lagrangian_modes}\end{eqnarray}
where $\phi_{L,k}$, $\phi_{R,k}$ refer to the individual normal modes of
the left and right resonator, respectively. The first two terms describe
two separate wave guides, while
the last term characterizes the coupling between both.

To transform
to the Hamiltonian, we consider the canonically conjugated momentum
$\pi_{L[R],k}=\partial L/\partial\dot{\phi}_{L[R],k}=c\dot{\phi}_{L[R],k}\mp2\frac{G}{d}(\sum_{k}\dot{\phi}_{R,k}-\sum_{k}\dot{\phi}_{L,k})$.
In the following we will restrict to a single mode in each resonator
($k=1$), and drop the label referring to the mode index. 
For $G/d\ll c$ (see
discussion below), we can simplify the expression for $\pi_{L[R],k}$ and 
consider $\pi_{L[R],k}=c\dot{\phi}_{L[R],k}$.
The first
two terms of (\ref{eq:Lagrangian_modes}) transform into two harmonic
oscillators of frequency $\omega_{L}$ and $\omega_{R}$. For
the coupling we have to consider $(\dot{\phi}_{R}-\dot{\phi}_{L})^{2}$
with $\dot{\phi}_{L[R]}=\pi_{L[R]}/c=i\sqrt{\frac{\hbar\omega_{L[R]}}{2c}}\left(a_{L[R]}^{\dagger}-a_{L[R]}\right)$,
see \cite{Blais2004Cavity-quantum-}. Using rotating wave approximation,
we find the Hamiltonian,
\begin{eqnarray*}
\mathcal{H} & = & \hbar\omega_{L}\left(1-\frac{G}{dc}\right)a_{L}^{\dagger}a_{L}+\hbar\omega_{R}\left(1-\frac{G}{dc}\right)a_{R}^{\dagger}a_{R}\\
 & + & \hbar\sqrt{\omega_{L}\omega_{R}}\frac{G}{dc}\left(a_{L}^{\dagger}a_{R}+a_{R}^{\dagger}a_{L}\right),
 \end{eqnarray*}
where we neglected the vacuum energy.
Note that none of the resonators in Fig.~\ref{fig:Setup_linear}
is short-circuited such that there are voltage antinodes at both ends
of each resonator allowing to have maximal coupling between $a_{L}$
and $a_{R}$, as well as to the feed and transmission line.

The coupling between the resonators has two effects. First,
both frequencies $\omega_{L}$, $\omega_{R}$, originally defined
for uncoupled modes, are lowered by a constant value.
In the following, this shift of frequency is neglected by simply redefining
the resonators' frequencies. More important is the coupling between
modes in terms of the coupling frequency\begin{eqnarray}
g & = & \sqrt{\omega_{L}\omega_{R}}\frac{G}{dc}.\label{eq:Coupling_frequency}\end{eqnarray}

Finally we take into account the motion of the nanomechanical beam
changing the left mode's bare eigenfrequency $\omega_{0}$ in the
way discussed above, $\omega_{L}(x)=\omega_{0}-\epsilon x$ (see Eq.~(\ref{eq:def_epsilon})
with $k=1$). The final Hamiltonian for the system depicted in Fig.~\ref{fig:Setup_linear}
then reads \begin{eqnarray}
\mathcal{H} & = & \hbar\left(\omega_{0}-\epsilon x\right)a_{L}^{\dagger}a_{L}+\hbar\omega_{0}a_{R}^{\dagger}a_{R}\nonumber \\
 & + & \hbar g\left(a_{L}^{\dagger}a_{R}+a_{R}^{\dagger}a_{L}\right).\label{eq:Final Hamiltonian_linear}\end{eqnarray}
In principle, according to (\ref{eq:Coupling_frequency}) with $\omega_{L}(x)$,
the coupling frequency $g$ depends on displacement $x$.
However, for typical parameters the dependence is negligible and
$g$ can be considered to be constant. The resonance frequency of (\ref{eq:Final Hamiltonian_linear})
is depicted in Fig.~\ref{fig:Setup_linear}d.

\textit{Coupling frequency comparable to the mechanical frequency
($g\simeq\Omega$).} - Given Eq. (\ref{eq:Coupling_frequency}), the
coupling frequency between the two resonator modes reads
$g/\omega_{0}=(c_{g}/c)\cdot(d_{g}/d)$,
where we defined the coupling line capacitance $c_{g}=G/d_{g}$ along
the length of the coupling region $d_{g}$.
In general, $c_{g}$ will
be much smaller than the line capacitance between each central
conductor and the ground plane $c$: first of all, the distance between
the two central conductors is significantly larger than the distance
between a single conductor and the adjacent ground plane. Second,
the capacitance between the central strip lines is shielded by the
grounded region in between. Here we crudely assume $c_{g}/c=10^{-2}$.
For $d$ in the cm range and $d_{g}\simeq0.1\text{mm}$
($d_{g}$ is chosen such that a several $10\,\mu\text{m}$ long nanomechanical
beam can be fabricated in between the region where the resonators
align), we have $d_{g}/d=10^{-2}$ and
the coupling between modes is $g/\omega_{0}=10^{-4}$ where $\omega_{0}$
will be in the GHz range. Common eigenfrequencies of nanomechanical
beams are $\Omega = 100\text{ \text{kHz}}$ - $10\text{ MHz}$. Hence, due
to their much smaller photon frequency, coupled multimode optomechanical
systems in the microwave regime naturally possess coupling frequencies
in the range of typical mechanical frequencies ($g\simeq\Omega$).
The relevance of this regime for instance to realize all kinds of
driven two- and multi-level photon dynamics in optomechanical systems has
been pointed out in \cite{Heinrich2010Photon-shuttle:}.

\textit{Transmission spectrum.} - As an example to emphasize the characteristics
of coupled optomechanical systems in the microwave regime and to demonstrate
implications of $g\simeq\Omega$ even in the presently accessible
regime of classic mechanical motion, we will discuss how the microwave field
in the setup of Fig.~\ref{fig:Setup_linear} can be manipulated in
terms of mechanical driving (see \cite{Unterreithmeier2009Universal-trans}
for a universal mechanical actuation scheme). Experimentally,
the impact can be most easily observed in terms of the transmission
spectrum. We assume the left resonator $a_{L}$ to be driven at frequency
$\omega_{L}$ via the feed line, while the transmission down the transmission
line is recorded. We consider the coupling of the left (right) resonator
to the feed (transmission) line in terms of the the resonators' decay rate $\kappa$.
Given the Hamiltonian~(\ref{eq:Final Hamiltonian_linear}),
using input/output theory, the equation of motion for the averaged
fields $\alpha_{L}=\langle a_{L}\rangle$, $\alpha_{R}=\langle a_{R}\rangle$
read \begin{eqnarray}
\frac{d}{dt}\alpha_{L} & = & \frac{1}{i}\left(-\epsilon x(t)\alpha_{L}+g\alpha_{R}\right)-\frac{\kappa}{2}\alpha_{L}-\sqrt{\kappa}b_{L}^{in}(t)\nonumber \\
\frac{d}{dt}\alpha_{R} & = & \frac{1}{i}g\alpha_{L}-\frac{\kappa}{2}\alpha_{R},\label{eq:EOM_average_fields}\end{eqnarray}
where $b_{L}^{in}(t)=e^{-i\Delta_{L}t}b^{in}$ describes the electromagnetic
drive along the feed line with amplitude $b^{in}$ and frequency $\omega_{L}$.
Here we used a rotating frame with laser detuning from resonance
$\Delta_{L}=\omega_{L}-\omega_{0}$. The transmission $T(t)=\kappa\langle a_{R}^{\dagger}(t)a_{R}(t)\rangle/\left(b^{in}\right)^{2}$
can be expressed as\begin{equation}
T(t)=\kappa^{2}\left|\int_{-\infty}^{t}G(t,t')e^{-i\Delta_{L}t'-(\kappa/2)(t-t')}\, dt'\right|^{2},\label{eq:Transmission_with_G}\end{equation}
where the phase comprises the feed line's drive and resonators' decay,
while the Green's function $G(t,t')$ describes the amplitude for
a photon to enter the left resonator's mode $a_{L}$ at time $t'$
and to be found in the right one $a_{R}$ later at time~$t$.

\begin{figure}
\includegraphics[width=1\columnwidth]{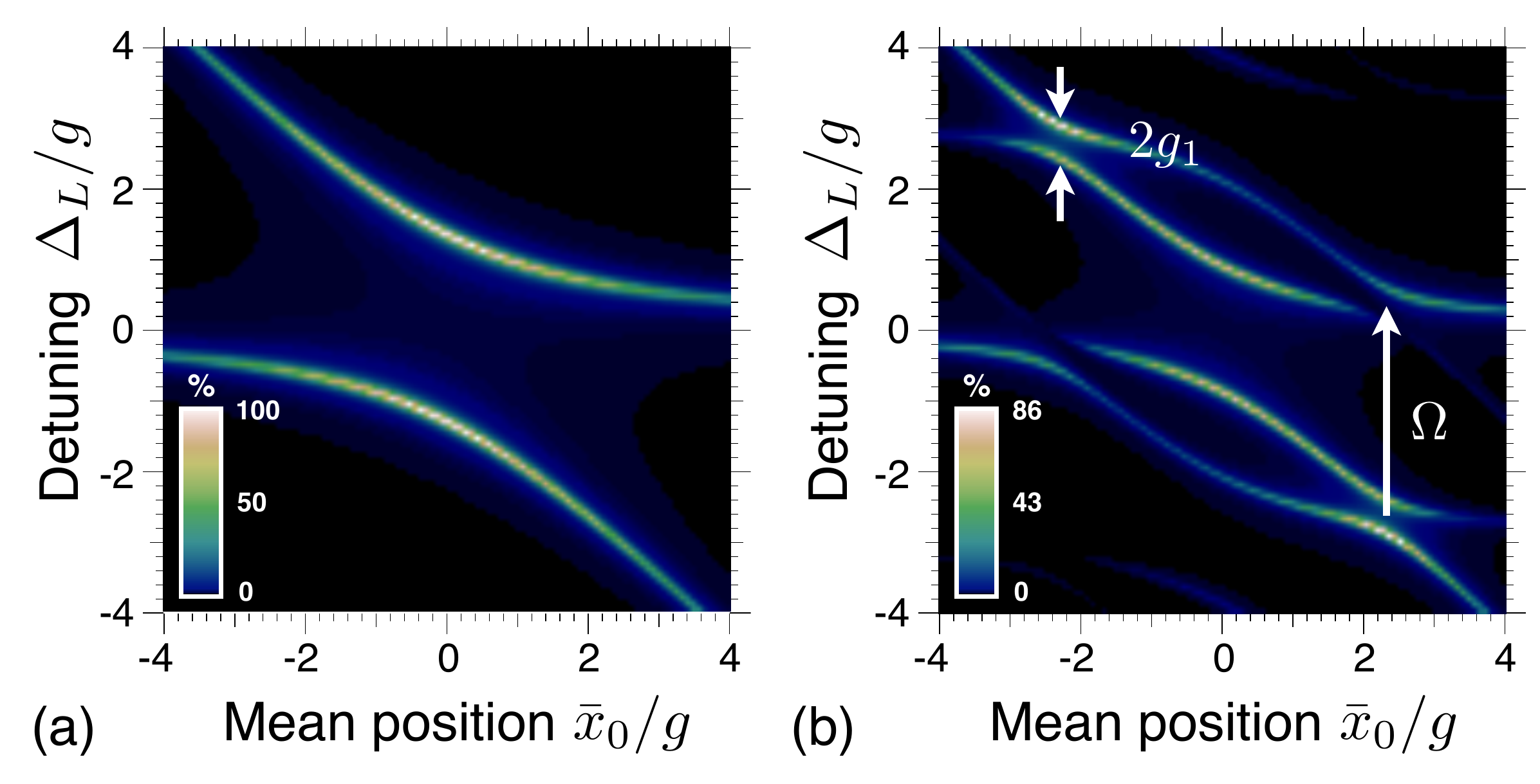}

\caption{Transmission spectrum for the setup depicted in Fig.~\ref{fig:Setup_linear}
for resonators' decay rate $\kappa=0.1g$: density plot for the time-averaged
transmission depending on mean mechanical displacement $\bar{x}_{0}=\epsilon x_{0}$,
and frequency detuning $\Delta_{L}=\omega_{L}-\omega_{0}$ of the
feed line's microwave drive at $\omega_{L}$ ($\omega_{0}$ denotes
the left mode's bare frequency for $x=0$). (a) Without mechanical
drive ($x(t)=x_{0}$); the spectrum is given by the resonance frequency
depicted in Fig.~\ref{fig:Setup_linear}d. (b) For mechanical driving
($x(t)=A\cos(\Omega t)+x_{0}$) with amplitude $\bar{A}=\epsilon A=1.5\Omega$
and frequency $\Omega=3g$; mechanical sidebands displaced by $\Omega$
appear and intersect the original photon branches where, due to mechanically
driven Rabi dynamics, high transmission and additional anticrossings
arise. The gap $2g_{1}$ is determined by the Bessel function $J_{1}$
according to $2g_{1}=2gJ_{1}(\bar{A}/\Omega)$. \label{fig:Transmission-spectum}}

\end{figure}

We take into account two scenarios: first, the beam is at rest given
a constant displacement $x(t)=x_{0}$; second, the beam is mechanically
driven to oscillate with amplitude $A$ and frequency $\Omega$ around
the mean position $x_{0}$, $x(t)=x_{0}+A\cos(\Omega t)$. Fig.~\ref{fig:Transmission-spectum}a
shows numerical results of the transmission spectrum without mechanical
driving. The spectrum corresponds to the system's resonance frequency
depicted in Fig.~\ref{fig:Setup_linear}d where the resonance width
is set by the resonators' decay rate $\kappa$.%
In contrast, Fig.~\ref{fig:Transmission-spectum}b shows the transmission
including mechanical driving with $\Omega=3g$, i.e. $g\simeq\Omega$
being characteristic for coupled microwave optomechanics.

To understand the main features of Fig.~\ref{fig:Transmission-spectum}b
we note that, in general, two processes are involved to observe transmission,
see (\ref{eq:Transmission_with_G}): first, the left resonator $a_{L}$
must be excited by the electromagnetic drive $b_{L}^{in}(t)=e^{-i\Delta_{L}t}b^{in}$;
second, the internal dynamics must be able to transfer photons from
$a_{L}$ to $a_{R}$. From (\ref{eq:EOM_average_fields}) the solution
$G(t,t')$ can be found to be\begin{equation}
G(t,t')=\tilde{\alpha}_{R}(t,t')e^{-i\phi(t')}\label{eq:Solution_Greens}\end{equation}
where $\phi(t')=(\bar{A}/\Omega)\sin(\Omega t')$ and $\tilde{\alpha}_{R}(t,t')$
is a solution to the driven two state problem\begin{equation}
\frac{d}{dt}\left(\begin{array}{c}
\tilde{\alpha}_{L}\\
\tilde{\alpha}_{R}\end{array}\right)=\frac{1}{i}\left(\begin{matrix}-\bar{x}_{0} & ge^{-i\phi(t)}\\
ge^{+i\phi(t)} & 0\end{matrix}\right)\left(\begin{array}{c}
\tilde{\alpha}_{L}\\
\tilde{\alpha}_{R}\end{array}\right),\label{eq:EOM_TLS_dyn}\end{equation}
with $t\geq t'$ and initial condition $\tilde{\alpha}_{L}(t',t')=1$,
$\tilde{\alpha}_{R}(t',t')=0$. Note that we expressed displacement
in terms of frequency; $\bar{A}=\epsilon A$, $\bar{x}_{0}=\epsilon x_{0}$.
For $\bar{A}\neq0$, in addition to the electromagnetic drive (see
$e^{-i\Delta t'}$ in (\ref{eq:Transmission_with_G})), the mechanical
driving can excite $a_{L}$ in terms of multiples of the mechanical
frequency $m\Omega$. This mechanical excitation is described
by the phase factor $e^{-i\phi(t')}=\sum_{m}J_{m}(\bar{A}/\Omega)e^{-im\Omega t'}$
in (\ref{eq:Solution_Greens}) and leads to mechanical sidebands in
the spectrum {[}cf. Fig.~\ref{fig:Transmission-spectum}b{]}. Note that
the individual process $m\Omega$ is described by a Bessel function
$J_{m}(\bar{A}/\Omega)$ and can be tuned by the driving strength.
Beyond the modified excitation, the driving significantly
changes the internal dynamics of the microwave fields, see Eq.~(\ref{eq:EOM_TLS_dyn}).
In particular the mechanical motion can initiate mechanically driven
Rabi dynamics exchanging photons between $a_{L}$ and $a_{R}$ that
leads to high transmission if the mechanical drive at $\Omega$ is
in resonance with the modes' frequency difference. For sufficiently
strong driving, the mechanically assisted process leads to additional
anticrossings in the spectrum resembling Autler-Townes splittings
known from quantum optics (see marker in Fig.~\ref{fig:Transmission-spectum}b).
From Eq. (\ref{eq:EOM_TLS_dyn}) we find that the spacing of this
first additional splitting scales according to $2gJ_{1}(\bar{A}/\Omega)$
and can likewise be tuned by the mechanical driving strength. All
this illustrates how, due to $g\simeq\Omega$, the microwave field
can extensively be manipulated by mechanical motion in terms of mechanically
driven coherent photon dynamics.

\begin{figure}
\includegraphics[width=1\columnwidth]{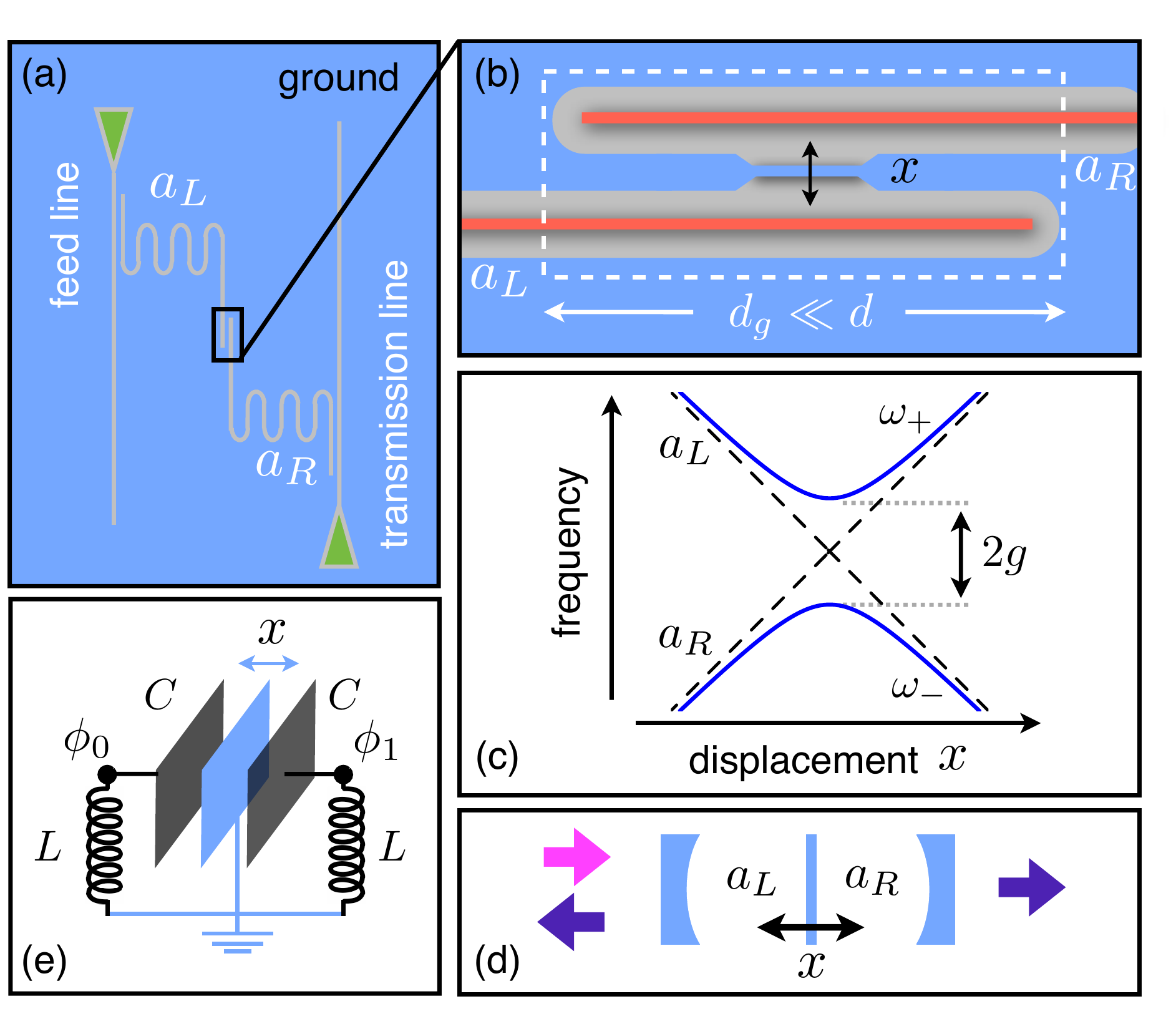}

\caption{\label{fig:Setup-quadratic}Schematic device geometry for two
microwave resonators $a_{L}$, $a_{R}$ with a nanomechanical device
coupled to \textit{both} of them. (a) Two stripline resonators (each of length $d$)
are coupled to external feed and transmission lines (green). The central conductors
of $a_{L}$, $a_{R}$ (red) are capacitively coupled in a small region
of length $d_{g}$ where the wave guides adjoin. (b) Between the two resonators
a small mechanical beam, connected to ground (blue), is placed. Its
displacement $x$ affects the line capacitance of both, $a_{L}$ and
$a_{R}$. (c) System's resonance frequency as function of displacement:
the beam's displacement linearly changes the bare modes' frequency
of $a_{L}$ and $a_{R}$ (dashed). Due to the coupling $g$ between
the resonators, there is an avoided crossing $2g$ in the eigenfrequencies
$\omega_{\pm}$ (blue). (d) Analogous optical setup with a movable
dielectric membrane placed in the middle of a cavity \cite{ThompsonStrong-dispersi}.
(e) Schematic realization with two microwave LC circuits where a
central plate is grounded and resonates against two others that build the LC circuits.
In the notation of Fig.~\ref{fig:Circuit-diagram} we get for the coupling
frequency $g=\omega_{LC}G/C$ with $\omega_{LC}=1/\sqrt{LC}$.} 

\end{figure}

\textbf{Coupling to the square of displacement.} - We present a modified
scheme comprising coupled microwave resonators that allows to couple
the photon number to the \textit{square} of mechanical displacement
[Fig.~\ref{fig:Setup-quadratic}a-b]. %
In contrast to the setup in Fig.~\ref{fig:Setup_linear}, here the nanomechanical beam is placed
in the region \textit{between} the two resonators, such that its motion
affects both simultaneously {[}Fig.~\ref{fig:Setup-quadratic}b{]}.
While for a given displacement $x$, the line capacitance $c$ of
the first resonator is increased, the one of the second wave
guide is decreased and vice versa.
According to our previous results, using the notation from above,
the Hamiltonian for this setup reads\begin{eqnarray}
H & = & \hbar\left(\omega_{0}-\epsilon x\right)a_{L}^{\dagger}a_{L}+\hbar\left(\omega_{0}+\epsilon x\right)a_{R}^{\dagger}a_{R}\nonumber \\
 & + & \hbar g\left(a_{L}^{\dagger}a_{R}+a_{R}^{\dagger}a_{L}\right),\label{eq:Hamiltonian_quadratic}\end{eqnarray}
where $g=\omega_{0}c_g d_g/cd$, see (\ref{eq:Coupling_frequency}) and (\ref{eq:Final Hamiltonian_linear}).

Fig.~\ref{fig:Setup-quadratic}c illustrates
the system's resonance frequency $\omega_{\pm}(x)=\omega_0\pm\sqrt{g^{2}+(\epsilon x)^{2}}$
as function of displacement. Naturally all the characteristics
of coupled multimode optomechanics in the microwave regime,
that have been discussed above, apply. In particular the hyperbola-shaped avoided
level crossing allows to realize Landau-Zener transitions and the dynamics of
Landau-Zener-Stueckelberg oscillations in the light field of a microwave setup.
At the extrema of $\omega_{\pm}(x)$ the
Hamiltonian allows an exclusive coupling to the square of
mechanical displacement $x^2$. Thus, in the following, we investigate the prospects to perform
QND Fock state detection using this microwave setting.
\begin{table*}
\begin{tabular*}{\textwidth}{@{\extracolsep{\fill}}|r|r|r|r|r|r|r|r|r|r|r|}
\hline 
$\epsilon$ {[}MHz/nm{]} & $m$ {[}pg{]} & $\Omega/2\pi$ {[}MHz{]} & $Q/10^{5}$ & $\kappa/\Omega$ & $g/2\pi$ {[}MHz{]} & $P_{in}$ {[}pW{]} & $x_{0}$ {[}pm{]} & $n_{add}$ & $T$ {[}mK{]} & $\Sigma$\tabularnewline
\hline
\hline 
$65$ & $10$ & $11$ & 3.5 & $1/70$ & 0.5 & 200 & 0.5 & 1 & 20 & 1.0\tabularnewline
\hline 
70 & 10 & 11 & 5.0 & 1/100 & 0.5 & 50 & 0.5 & 1 & 20 & 1.2\tabularnewline
\hline
\end{tabular*}

\caption{\label{tab:qjump_parameters}Two sets of experimental parameters that
would allow to observe an individual quantum jump from the mechanical
ground state to the first excited state, with a signal-to-noise ratio $\Sigma\geq1$. Further parameter
$\omega_{c}/2\pi=5\text{ GHz}$.}

\end{table*}

\textit{Fock state detection.} - In principle, an exclusive coupling of the photon number to $x^{2}$
allows to perform QND Fock state detection and to observe
quantum jumps of a mechanical resonator \cite{Braginsky1980Quantum-Nondemo,Braginsky1992Quantum-Measure}. Indeed the Hamiltonian~(\ref{eq:Hamiltonian_quadratic}) corresponds
to the one found for an optical setup (see Fig.~\ref{fig:Setup-quadratic}d), that
generated a lot of interest in this regard \cite{ThompsonStrong-dispersi,Jayich2008Dispersive-opto}.
We focus on one microwave mode with annihilation operator $a$
and expand the resonance frequency $\omega_{+}(x)$ around $x_{0}$.
For $x_{0}=0$ the linear contribution vanishes and the Hamiltonian
reads $H=\hbar\left(\omega_{+}(0)+\frac{1}{2}\omega_{+}^{''}(0)x_{zp}^{2}[b^{\dagger}+b]^{2}\right)a^{\dagger}a+\hbar\Omega b^{\dagger}b$,
where we considered the phonon number operator $n=b^{\dagger}b$ and
quantized $(x-x_{0})$ using the mechanical beam's displacement operator
$\hat{x}=x_{zp}\left(b^{\dagger}+b\right)$. The zero-point displacement
$x_{zp}=\sqrt{\hbar/2m\Omega}$ is determined by the mechanical mass
$m$ and frequency $\Omega$. Applying rotating wave approximation
(RWA), $\left(b^{\dagger}+b\right)^{2}\simeq2n+1$, we immediately see that $\left[H,n\right]=0$.
For a potential experiment we consider a scheme in analogy to the
one proposed for the optical setup \cite{ThompsonStrong-dispersi}. The mechanics is
cooled to the quantum mechanical ground state \cite{Teufel2008Dynamical-Backa, Rocheleau2010Preparation-and}.
After switching off the cooling, the phonon number $n$ is measured
via the frequency of the microwave mode. To detect a quantum jump
from $n=0$ to $n=1$, the frequency
shift per phonon $\Delta\omega=\omega_{+}^{''}(0)x_{zp}^{2}$, where
$\omega_{+}^{''}(0)=\epsilon^{2}/g$, must be resolved within the
lifetime of the phonon ground state $\tau^{(0)}$. Given the imprecision
of the frequency measurement in terms of the angular frequency noise power
spectral density $S_{\omega\omega}$ (in units $\text{s}^{-2}/\text{Hz}$),
the signal-to-noise ratio reads $\Sigma=(\Delta\omega)^{2}\tau^{(0)}/S_{\omega\omega}$
\cite{ThompsonStrong-dispersi}. In contrast to the optical regime, where shot-noise-limited
frequency measurements are routinely achieved, microwave setups
in general suffer from amplifier noise adding $n_{add}$ quanta of
noise beyond the shot-noise limit in a Pound-Drever-Hall scheme,
such that $S_{\omega\omega}=(n_{add}+1/2)\kappa^{2}\hbar\omega_{c}/16P_{in}$
(see \cite{Black2001An-introduction}). $P_{in}$ denotes the incident power and $\omega_{c}$
is the resonance frequency of the cavity. While commercially available
systems add a significant amount of noise, a new Josephson parametric
amplifier achieved $n_{add}<1/2$ \cite{Castellanos-Beltran2008Amplification-a}. This technique
has already been used for displacement measurements in a microwave
optomechanical system with $n_{add}=1.3$ \cite{Teufel2009Nanomechanical-}. Essentially,
the total lifetime $\tau^{(0)}=1/(\tau_{T}^{-1}+\tau_{RWA}^{-1}+\tau_{lin}^{-1})$
is set by the thermal lifetime $\tau_{T}=\hbar Q/k_{B}T$ via the
mechanical quality factor $Q$ and the chip temperature $T$.
Additional contributions due to the RWA ($\tau_{RWA}$) and imprecise
positioning $x_{0}\neq0$ ($\tau_{lin}$) will be determined via Fermi's
golden rule rates (see \cite{ThompsonStrong-dispersi}).

For microwave setups using a small nanomechanical beam manufactured
close to the central conductor of a stripline resonator, achieving
optomechanical couplings of $\epsilon\simeq1-100\text{kHz/nm}$ 
\cite{Regal2008Measuring-nanom,Teufel2008Dynamical-Backa,Rocheleau2010Preparation-and},
the frequency shift per phonon $\Delta\omega\propto\epsilon^{2}$
turns out to be extremely small making Fock state detection impossible.
A new on-chip microwave system however, consisting of an LC circuit
where the plates of a parallel-plate condensator mechanically resonate,
achieves $\epsilon=65\text{MHz/nm}$ \cite{JohnTeufel_PrivCom}. Our proposal
transfers to this scheme by stacking three such plates, see Fig.~\ref{fig:Setup-quadratic}e. For
experimentally realistic parameters \cite{JohnTeufel_PrivCom}, a calculation of $\Sigma$ yields that
a setup with this optomechanical coupling would allow to detect an individual quantum jump
from the mechanical ground state to the
first exited state, see Tab.~\ref{tab:qjump_parameters}.
Note that, in contrast to the setup discussed in \cite{ThompsonStrong-dispersi}, the parameters
here are already in the small-splitting regime $g<\Omega$,
and the details of Fock state detection in that regime may require further analysis.
Finally, we point out that such a setup, even for $\Sigma < 1$,
would allow to measure {}``phonon shot noise'',
i.e. quantum energy fluctuations around an average phonon number, of a mechanically
driven, ground-state-cooled mechanical oscillator \cite{Clerk2010Quantum-Measure}.

\textbf{Conclusion.} - To conclude, we introduced and analyzed
theoretically coupled multimode
optomechanical systems for the microwave regime.
In contrast to the optical domain, these systems possess coupling
frequencies between the electromagnetic modes that are naturally
in the range of typical mechanical frequencies ($g\simeq\Omega$).
By calculating the transmission spectrum, we demonstrated how this
allows to manipulate the microwave field dynamics in terms of mechanical
driving. In principle $g\simeq\Omega$ enables to realize all kinds
of driven two- and multi-level dynamics known from quantum optics in the
microwave light field.
Our discussion mostly focussed on classical mechanical motion.
However, for mechanical oscillators in
the quantum regime, coupled multimode systems with $g\simeq\Omega$
will be particularly interesting. For instance it might be possible
to realize hybridized states which are superpositions of states with
a photon being in different modes and phonons in the mechanics.
We furthermore proposed a multimode setup that allows to couple the microwave 
photon number to the square of mechanical displacement and enables
QND Fock state detection. For experimentally realistic parameters we predicted
the possibility to detect an individual quantum jump from the mechanical ground state
to the first excited state. The same scheme also allows to measure phonon shot noise.
Both experiments would constitute a major breakthrough.

\acknowledgments
We acknowledge fruitful discussions with Rudolf Gross, Eva Weig, John Teufel and
Konrad Lehnert as well as support by the DFG (NIM, SFB 631, Emmy-Noether
program), GIF and DIP.

\bibliography{Literature.bib}

\end{document}